\documentclass[preprintnumbers,article,amsmath,amssymb,floatfix,10pt,prd,twocolumn,
superscriptaddress,nofootinbib]{revtex4-2}
\usepackage{bm}
\usepackage{amsfonts}
\usepackage{latexsym}
\usepackage[latin1]{inputenc}
\usepackage{graphicx}
\usepackage{amsmath}
\usepackage{palatino}
\usepackage{mathpazo}
\usepackage{textcomp}
\usepackage{float}
\usepackage{booktabs}
\usepackage{dcolumn}
\usepackage{ragged2e}
\usepackage{hyperref}
\hypersetup{colorlinks,citecolor=blue}
\hypersetup{colorlinks=true,linkcolor=red,filecolor=magenta,    urlcolor=blue}
\usepackage{amsmath}
\usepackage{xcolor}
\usepackage{orcidlink}
\usepackage{epsfig}
\usepackage{caption}
\usepackage{subcaption}
\usepackage{commath}
\def\jnl@style{\it}
\def\aaref@jnl#1{{\jnl@style#1}}

\def\aaref@jnl#1{{\jnl@style#1}}

\def\aj{\aaref@jnl{AJ}}                   
\def\apj{\aaref@jnl{ApJ}}                 
\def\apjl{\aaref@jnl{ApJ}}                
\def\apjs{\aaref@jnl{ApJS}}               
\def\apss{\aaref@jnl{Ap\&SS}}             
\def\aap{\aaref@jnl{A\&A}}                
\def\aapr{\aaref@jnl{A\&A~Rev.}}          
\def\aaps{\aaref@jnl{A\&AS}}              
\def\mnras{\aaref@jnl{Mon.~Not.~Roy.~Astron.~Soc.}}             
\def\prd{\aaref@jnl{Phys.~Rev.~D}}        
\def\prc{\aaref@jnl{Phys.~Rev.~C}}  
\def\prl{\aaref@jnl{Phys.~Rev.~Lett.}}    
\def\qjras{\aaref@jnl{QJRAS}}             
\def\skytel{\aaref@jnl{S\&T}}             
\def\ssr{\aaref@jnl{Space~Sci.~Rev.}}     
\def\zap{\aaref@jnl{ZAp}}                 
\def\nat{\aaref@jnl{Nature}}              
\def\aplett{\aaref@jnl{Astrophys.~Lett.}} 
\def\apspr{\aaref@jnl{Astrophys.~Space~Phys.~Res.}} 
\def\physrep{\aaref@jnl{Phys.~Rep.}}      
\def\physscr{\aaref@jnl{Phys.~Scr}}       
\def\commat{\aaref@jnl{Comm.~Math.~Phys.}}              
\def\science{\aaref@jnl{Science}}               
\def\cqg{\aaref@jnl{Classical Quant.~Grav.}}            
\def\jpcs{\aaref@jnl{JPCS}}                                     
\def\ijmpd{\aaref@jnl{Int.~J.~Mod.~Phys.~D}}                    
\def\grg{\aaref@jnl{Gen.~Relat.~Gravit.}}               
\def\rpp{\aaref@jnl{Rep.~Prog.~Phys.}}          
\def\npa{\aaref@jnl{Nucl.~Phys.~A}}        
\def\lrr{\aaref@jnl{Living Rev.~Rel.}}                   
\def\jcap{\aaref@jnl{J.~Cosmology Astropart.~Phys.}}    
\def\rmp{\aaref@jnl{Rev.~Mod.~Phys.}}   
\def\epjc{\aaref@jnl{Eur.~Phys.~J.~C}} 
\def\plb{\aaref@jnl{~Phy.~Lett.~B}} 
\def\mpla{\aaref@jnl{Mod.~Phy.~Lett.~A}} 
\def\arxiv{\aaref@jnl{arxiv.org}}

\allowdisplaybreaks[1]

\begin{document}
\color{black}       
\title{Interaction of divergence-free deceleration parameter in Weyl-type $f(Q,T)$ gravity}

\author{Gaurav N. Gadbail\orcidlink{0000-0003-0684-9702}}
\email{gauravgadbail6@gmail.com}
\affiliation{Department of Mathematics, Birla Institute of Technology and
Science-Pilani,\\ Hyderabad Campus, Hyderabad-500078, India.}

\author{Simran Arora\orcidlink{0000-0003-0326-8945}}
\email{dawrasimran27@gmail.com}
\affiliation{Department of Mathematics, Birla Institute of Technology and
Science-Pilani,\\ Hyderabad Campus, Hyderabad-500078, India.}

\author{Praveen Kumar\orcidlink{0000-0002-8201-6019}}
\email{pkumar6743@gmail.com}
\affiliation{Department of Mathematics, G H Raisoni College of Engineering, Nagpur-440016, India.}

\author{P.K. Sahoo\orcidlink{0000-0003-2130-8832}}
\email{Corresponding author: pksahoo@hyderabad.bits-pilani.ac.in}
\affiliation{Department of Mathematics, Birla Institute of Technology and
Science-Pilani,\\ Hyderabad Campus, Hyderabad-500078, India.}

%

\begin{abstract}
\textbf{Abstract:} We study an extension of symmetric teleparallel gravity i.e. Weyl-type $f(Q,T)$ gravity and the divergence-free parametrization of the deceleration parameter $q(z) = q_{0}+q_{1}\frac{z(1+z)}{1+z^2}$ ($q_{0}$ and $q_{1}$ are free constants) to explore the evolution of the universe. By considering the above parametric form of $q$, we derive the Hubble solution and further impose it in the Friedmann equations of Weyl-type $f(Q, T)$ gravity. To see whether this model can challenge the $\Lambda$CDM limits, we computed the constraints on the model parameters using the Bayesian analysis for the Observational Hubble data ($OHD$) and the Pantheon sample ($SNe\,Ia$). Furthermore, the deceleration parameter depicts the accelerating behavior of the universe with the present value $q_0$ and the transition redshift $z_t$ (at which the expansion transits from deceleration to acceleration) with $1-\sigma$ and $2-\sigma$ confidence level. We also examine the evolution of the energy density, pressure, and effective equation of state parameters. Finally, we demonstrate that the divergence-free parametric form of the deceleration parameter is consistent with the Weyl-type $f(Q,T)$ gravity.\\

\end{abstract}
\date{\today}
\maketitle

\section{Introduction}
The most significant challenge theoretical physics has faced is the late-time acceleration expansion of the universe. In accordance with different observational evidence, such as Type $Ia$ supernovae $(SNe\,Ia)$ \cite{Perlmutter/1999,Riess/1998,Riess/2004}, cosmic microwave background radiation (CMBR) \cite{Spergel/2007}, and several large-scale structural measurements \cite{Koivisto/2006,Daniel/2008}, it is now widely believed that the universe is undergoing a phase of accelerated expansion. The most popular explanation is that dark energy now dominates the universe.  Nevertheless, none of the current dark energy models is entirely acceptable. The most competitive cosmological dark energy model is a $\Lambda$CDM model, often known as the cosmological constant model with the equation of state $\omega_{\Lambda}=-1$.
The cosmological constant DE has proven to be a valuable model to account for and explain many cosmological data. Some theoretical arguments suggest that it could be necessary to consider dynamical dark energy. In particular, the coincidence problem- a situation where the cosmological constant dominates the dynamics of the universe at late times, cannot be explained by any fundamental hypothesis. Another significant issue is the "fine-tuning problem", which refers to the disparity between the observed value and the theoretically predicted value of the cosmological constant \cite{Hinshaw/2013,Zhao/2012}. Numerous theories for dark energy with a time-evolving energy density have also been proposed in the literature to solve or at least lessen the aforementioned cosmological challenge \cite{Ratra/1988,Caldwell/1998,Caldwell/2002,Caldwell/2003,Feng/2005,Huang/2005,Afshordi/2007}.\\
Another way to explain the present acceleration of the universe is to modify the geometry of spacetime. We can do this by modifying the Einstein-Hilbert general relativity (GR) action. One of the best theories in this context is $f(R)$ gravity \cite{Capozziello/2008,Sharif/2013}, which extends the gravitational action of general relativity to include a generic scalar curvature function. Modified gravity based on a non-minimal interaction between matter and geometry, such as $f(R,T)$ theory \cite{Xing/2016, Harko/2014, Moraes/2017, Harko/2011,Yousaf/2016}, $f(R,G)$ theory \cite{Elizalde/2010, Bamba/2010}, are another extensions proposed in the literature.\\
The Levi-Civita connection is the affine connection specified in the standard description of General Relativity on the spacetime manifold, which utilizes Riemannian geometry. However, there are several choices for affine connections on any manifold and may result in various equivalent descriptions of gravity \cite{Heisenberg/2019,Harada/2020}, which may offer multiple perspectives on understanding. One can define the Teleparallel equivalent of GR (TEGR) by selecting a connection that relaxes the constraint on torsion while requiring both curvature and non-metricity to disappear \cite{Maluf/525}. Additionally, a new theory named Symmetric Teleparallel equivalent of GR (STEGR) has been proposed by choosing a connection that relaxes the constraint on non-metricity while necessitating the annihilation of both curvature and torsion \cite{Nester/1999,Adak/2006}. In an extension of STEGR, one may formulate the $f(Q)$ gravity \cite{Jimenez/2018}, where $Q$ is a non-metricity representing the fundamental geometrical variable characterizing how a vector's length changes when transported. Moreover, a recent extension of $f(Q)$ gravity known as the $f(Q,T)$ gravity \cite{Xu/2019} is based on the non-minimal coupling between the non-metricity $Q$ and the trace of the energy-momentum tensor $T$. The $f(Q,T)$ theory is constructed in a manner similar to the $f(R,T)$ theory. Here, the standard Ricci scalar $R$ is replaced by the non-metricity $Q$, representing the symmetric teleparallel gravity formulation.

Numerous studies have shown that $f(Q,T)$ gravity is useful in explaining the present cosmic acceleration of the universe and offering a feasible solution to the dark energy problem \cite{Arora/2020,Arora/2021}. To review $f(Q,T)$ gravity in different aspects, one can check references \cite{Bhattacharjee/2020,Gadbail/2022,Najera/2022}. Furthermore,  Yixin et al. \cite{Xu/2020} explored $f (Q, T)$ gravity in the content of proper Weyl geometry called Weyl-type $f(Q,T)$ gravity theory. However, in the present approach to $f(Q, T)$ type gravity theories, we have formulated the nonmetricity $Q$ employing the prescriptions of the Weyl geometry, in which the covariant divergence of the metric tensor is given by the product of the metric tensor and Weyl vector $w_{\mu}$. The scalar nonmetricity is related simply to the square of the Weyl vector as $Q =-6w^2$. Thus, the Weyl vector and the metric tensor describe all of the geometric properties of the theory. Yixin et al. \cite{Xu/2020} analyzed the cosmological implications of the Weyl-type $f(Q,T)$ theory for three classes of specific models. The resulting solutions represent both accelerating and decelerating evolutionary phases of the universe. It is seen that Weyl type $f(Q,T)$ gravity may be considered a possible candidate for characterizing the early and late phases of cosmic evolution. Yang et al. \cite{Yang/2021} also derived the geodesic and Raychaudhuri equations using the Weyl $f(Q,T)$ theory. To get in touch with works in Weyl-type $f(Q,T)$ gravity theory, one can refer \cite{Gadbail/2021,Gadbail/2021a}. \\
In this work, we consider the divergence-free parametric form of deceleration parameter, and investigate the FLRW universe in the Weyl-type $f(Q,T)$ gravity theory by a functional form $f(Q,T)$ as $f(Q,T)=\alpha Q+ \frac{\beta}{6\kappa^2}T$, where $\alpha$ and $\beta$ are free parameters. The present model is based on the divergence-free parametric form of the deceleration parameter to get the exact solutions to the field equations.
The present article is outlined as follows: In section \ref{section 2}, we present the Weyl Type $f(Q,T)$ gravity formalism. The solution of the field equation and the Hubble parameter is presented in section \ref{section 3}.  In section \ref{section 4}, we discuss the observational data and the methodology used to constrain the model parameters. The behavior of cosmological parameters, including the deceleration parameter, pressure, density, and  EoS parameter, is covered in section \ref{section 5}. Finally, in the last section \ref{section 6}, we briefly discuss our results.

\section{Field equations}
\label{section 2}

We start with the gravitational action in the Weyl-type $f(Q,T)$ given by: \cite{Xu/2020} 
\begin{multline}
\label{1}
 S=\int \left[ \kappa^2f(Q,T)-\frac{1}{4}W_{\alpha \beta}W^{\alpha \beta}-\frac{1}{2}m^2 w_\alpha w^\alpha+\right.\\ 
 \left. \lambda \tilde{R}+\mathcal{L}_m\right]\sqrt{-g}d^4x,
\end{multline}    
where, $\tilde{R}=(R+6\nabla_\mu w^\mu-6w_\mu w^\mu)$, $\kappa^2=\frac{1}{16\pi G}$, a particle associated with a vector field has a mass of $m$, $\mathcal{L}_m$ is the matter Lagrangian. Furthermore, $f$ can be defined as an arbitrary function of the non-metricity $Q$ and the trace of the matter-energy-momentum tensor $T$. Also, the ordinary kinetic term and the mass term of the vector field are represented by the second and third terms in the action, respectively.\\ 
We can define the non-metricity scalar as 
\begin{equation}
\label{2}
Q\equiv- g^{\alpha \beta}\left(L^\mu_{\nu\beta}L^\nu_{\beta\mu}-L^\mu_{\nu\mu}L^\nu_{\alpha \beta}\right),
\end{equation}
where, $L^\lambda_{\alpha \beta}$ is the deformation tensor
\begin{equation}
\label{3}
L^\lambda_{\alpha \beta}=-\frac{1}{2}g^{\lambda\gamma}\left(Q_{\alpha\gamma\beta}+Q_{\beta\gamma\alpha}-Q_{\gamma\alpha \beta}\right).
\end{equation}
One can relate that the Levi-Civita connection and the metric in Riemannian geometry, i.e., $\nabla_\mu g_{\alpha \beta}=0$ are both compatible. Whereas, this seems different in Weyl geometry by
\begin{equation}
\label{4}
Q_{\mu\alpha\beta}\equiv\widetilde{\nabla}_\mu g_{\alpha\beta}=\partial_\mu g_{\alpha\beta}-\widetilde{\Gamma}^\rho_{\mu \alpha}g_{\rho\beta}-\widetilde{\Gamma}^\rho_{\mu \beta}g_{\rho\alpha}=2w_\mu g_{\alpha\beta},
\end{equation}
where, $\widetilde{\Gamma}^\lambda_{\alpha\beta}\equiv\Gamma^\lambda_{\alpha\beta}+g_{\alpha\beta}w^\lambda-\delta^\lambda_\alpha w_\beta-\delta^\lambda_\beta w_\alpha$ and $\Gamma^\lambda_{\alpha\beta}$ is the Christoffel symbol with respect to the metric $g_{\alpha\beta}$.\\
Using Eqs. \eqref{2}-\eqref{4}, we obtain the following relation: 
\begin{equation}
\label{5}
Q=-6w^2,
\end{equation}
 with the effective dynamical mass of the vector field as $m^2_{eff}=m^2+12\kappa^2f_Q+12\lambda$. The variation principle with respect to the metric tensor and Weyl vector on \eqref{1} yields the field equation. 
\begin{multline}
\label{7}
\frac{1}{2}\left(T_{\alpha\beta}+S_{\alpha\beta}\right)-\kappa^2f_T\left(T_{\alpha\beta}+\Theta_{\alpha\beta}\right)=-\frac{\kappa^2}{2}g_{\alpha\beta}f\\
-6\kappa^2f_Q w_\alpha w_\beta +\lambda\left(R_{\alpha\beta}-6w_\alpha w_\beta +3g_{\alpha\beta}\nabla_\rho w^\rho \right)\\
+3g_{\alpha\beta}w^\rho \nabla_\rho \lambda 
-6w_{(\alpha}\nabla_{\beta )}\lambda+g_{\alpha\beta}\square \lambda-\nabla_\alpha\nabla_\beta \lambda,
\end{multline}
where
\begin{equation}
\label{8}
T_{\alpha\beta}\equiv-\frac{2}{\sqrt{-g}}\frac{\delta(\sqrt{-g}L_m)}{\delta g^{\alpha\beta}},
\end{equation} 
\begin{equation}
\label{9}
f_T\equiv \frac{\partial f(Q,T)}{\partial T},
f_Q\equiv\frac{\partial f(Q,T)}{\partial Q}.
\end{equation}
respectively. Also the expression for $\Theta_{\alpha\beta}$ is defined as
\begin{equation}
\label{10}
\Theta_{\alpha\beta}=g^{\mu\nu}\frac{\delta T_{\mu\nu}}{\delta g_{\alpha\beta}}=g_{\alpha\beta}L_m-2T_{\alpha\beta}-2g^{\mu\nu}\frac{\delta^2 L_m}{\delta g^{\alpha\beta}\delta g^{\mu\nu}}.
\end{equation}
Here, $S_{\alpha\beta}$ is the re-scaled energy momentum tensor of the free Proca field 
\begin{equation}
\label{11}
S_{\alpha\beta}=-\frac{1}{4}g_{\alpha\beta}W_{\rho\sigma}W^{\rho\sigma}+W_{\alpha\rho}W^\rho_\beta -\frac{1}{2}m^2g_{\alpha\beta}w_\rho w^\rho +m^2 w_\alpha w_\beta,
\end{equation}
with $W_{\alpha\beta}=\nabla_\beta w_\alpha-\nabla_\alpha w_\beta$ .\\

\section{Cosmological Model and Solutions}
\label{section 3}

Consider a flat Friedmann-Lemaitre-Robertson-Walker (FLRW) universe, which is represented by the isotropic, homogeneous, and spatially flat metric
\begin{equation}
\label{12}
ds^2=-dt^2+a^2(t)\delta_{ij}dx^i dx^j ,
\end{equation}
where, $a(t)$ is the scale factor. The vector field due to spatial symmetry is assumed to be of the form $w_\alpha=\left[\psi (t),0,0,0\right]$. Hence, we have $w^2=w_\alpha w^\alpha=-\psi^2(t)$ with 
 $Q=-6w^2=6\psi^2(t)$.\\
Here, we adapt the comoving coordinates $u^{\alpha}= \left(-1,0,0,0\right)$ along with $u^\alpha \nabla_\alpha=\frac{d}{dt}$ and $H=\frac{\dot{a}}{a}$. We also assume the Lagrangian of the perfect fluid as $\mathcal{L}_m=p$.\\

The corresponding energy momentum tensor for the perfect fluid $T_{\mu\nu}$ and $\Theta^\mu_\nu$ lead to $T^\mu_\nu=diag\left(-\rho,p,p,p\right)$ and $\Theta^\mu_\nu=\delta^\mu_\nu p-2T^\mu_\nu=diag\left(2\rho+p,-p,-p,-p\right)$, 
where $\rho$ and $p$ are the energy density and the pressure, respectively. 

The generalised Proca equation and the flat space restriction for the cosmological case read as
\begin{eqnarray}
\dot{\psi} &=&\dot{H}+2H^2+\psi^2-3H\psi,  \label{17}\\
\dot{\lambda}&=&\left(-\frac{1}{6}m^2-2\kappa^2f_Q-2\lambda\right)
\psi-\frac{1}{6}m^2_{eff}\psi , \label{18} \\ 
 \partial_i \lambda &=&0. \label{19}
\end{eqnarray}

 Imposing an FLRW metric in equation \eqref{7}, we get the two Friedmann equations as
\begin{multline}
\label{20}
\kappa^2f_T\left(\rho+p\right)+\frac{1}{2}\rho=\frac{\kappa^2}{2}f-\left(6\kappa^2f_Q+\frac{1}{4}m^2\right)\psi^2 \\
-3\lambda\left(\psi^2-H^2\right)-3\dot{\lambda}\left(\psi-H\right),
\end{multline}

\begin{multline}
\label{21}
-\frac{1}{2}p=\frac{\kappa^2}{2}f+\frac{m^2\psi^2}{4}+\lambda\left(3\psi^2+3H^2+2\dot{H}\right)\\
+\left(3\psi+2H\right)\dot{\lambda}+\ddot{\lambda}.
\end{multline}
where dot $(.)$ represents the time derivative and $f_Q$, and $f_T$ defines the derivative with respect to $Q$ and $T$, respectively.\\
In this work, we assume the functional form $f(Q,T)=\alpha Q+\frac{\beta}{6\kappa^2}T$, where $\alpha$ and $\beta$ are model parameters. It is important to note that $\beta=0$ and $\alpha=-1$ corresponds to $f(Q,T)= - Q$, i.e., a case of the successful General Relativity theory (GR). Furthermore, at $T=0$, the theory simplifies to $f(Q)=\alpha Q$ gravity, which is equivalent to GR and passes all Solar System tests. Further, using the relation $\nabla_{\lambda}g_{\mu\nu}=-w_{\lambda}g_{\mu\nu}$, we can obtain $\psi(t)=-6H(t)$, where $H(t)=\frac{\dot{a}}{a}$. For the choice of $f(Q,T)$ and solving equations \eqref{20} and \eqref{21}, we obtain the expression for pressure $p$ and energy density $\rho$ as follows:
\begin{equation}
\label{22}
p=-k_1\,H^2-k_2\,\dot{H},
\end{equation}
where 
\begin{align*}
k_1&=\left(36\left(\frac{18}{\beta+3}\left(\alpha+1\right)+\frac{3M^2}{2\left(\beta+3\right)}\right)+\frac{18}{2\beta+3}\right) , \\
k_2&=\frac{18\left(\beta+2\right)}{\left(2\beta+3\right)\left(\beta+3\right)}.
\end{align*}
and 
\begin{equation}
\label{23}
\rho=k_3\,H^2-k_4\,\dot{H}.
\end{equation}
where 
\begin{align*}
k_3&=\left(\frac{-9\left(11\beta+24\right)\left(24\alpha+25\right)}{4\left(\beta+2\right)\left(\beta+3\right)}+\frac{29\beta+72}{2\left(2\beta+3\right)\left(\beta+2\right)}\right), \\   
k_4& =\frac{9\beta}{2\left(2\beta+3\right)\left(\beta+3\right)}.\\
\end{align*}
The effective equation of state $\omega_{eff}=\frac{p}{\rho}$ becomes
\begin{equation}
\label{24}
\omega_{eff}=\frac{-k_1\,H^2-k_2\,\dot{H}}{k_3\,H^2-k_4\,\dot{H}}
\end{equation}
Considerable cosmological observations show that the expansion of the universe has accelerated after experiencing a deceleration. Since then, the deceleration parameter $q$ has been frequently utilized to characterize cosmic history, at least up until this point. In this sense, some studies used different parametric variants of $q$ while others explored non-parametric forms.
The literature has suggested several parametrization forms of $q$. However, in this work, we will examine the divergence-free parametric form of the deceleration \cite{Mamon/2016,Hanafy/2019} given below:
\begin{equation}
\label{25}
	q(z) = q_{0}+q_{1}\frac{z(1+z)}{1+z^2}.
\end{equation}
where $q_{0}$ is the present value of $q$ and $q_1$ indicates the variation of the deceleration parameter with respect to $z$.\\
Mathematically, this parametric form has two possible cases:
\begin{itemize}
    \item The function $q(z)$ reduces to $q(z)\sim q_{0}+ q_{1}$ at high redshift (i.e., $z>>1$).
    \item The function $q(z)$ reduces to $q(z)\sim q_{0}+ q_{1}z$ at low redshift (i.e., $z<<1$), (this model is not reliable at high redshift).
\end{itemize} 
Another motivation behind this parametrization is that it provides a finite value of $q$ throughout the entire range, $z\in [-1,\infty]$, and so encompasses the entire evolutionary history of the universe. It is important to note that the above parametric form of $q(z)$ was inspired by one of the most prominent divergence-free parametrizations of the dark energy equation of state \cite{Barboza/2008}, and it seemed to be sufficiently adaptable to match the $q(z)$ behavior of a large class of accelerating models.\\
The following equation relates the deceleration parameter and the Hubble parameter 
\begin{equation}
 \label{26}
 	H(z) = H_{0}\, exp\Big(\int_{0}^{z} \frac{1+ q(z)}{(1 + z)} dz \Big),
 \end{equation}
Inserting Eq. \eqref{25} into Eq. \eqref{26}, we get $H(z)$ as
\begin{equation}
\label{27}
	H(z) = H_{0} (1+z)^{(1+ q_{0})} (1+z^2)^{\frac{q_1}{2}}.
\end{equation}
The derivative of the Hubble parameter with respect to time can be written as 
\begin{equation}
\label{28}
\dot{H}=\frac{dH}{dt}=-(1+z)H(z)\frac{dH}{dz}
\end{equation}
Using equations \eqref{27} and \eqref{28}, the expressions for $p$ and $\rho$ as functions of $z$ are the following:
\begin{multline}
\label{29}
p=-H_0^2 (z+1)^{2q_0+2} \left(z^2+1\right)^{q_1-1}\\
 \left[k_1 \left(z^2+1\right)-k_2 \left(q_0 \left(z^2+1\right)+q_1 (z+1) z+z^2+1\right)\right],
\end{multline}
and
\begin{multline}
\label{30}
\rho=H_0^2 (z+1)^{2q_0+2} \left(z^2+1\right)^{q_1-1}\\
\left[k_3 \left(z^2+1\right)+k_4\left(q_0 \left(z^2+1\right)+q_1 (z+1) z+z^2+1\right)\right].
\end{multline}

\section{observational data}
\label{section 4}
Now, we test the viability of the model using the recent observational data, namely, the observational Hubble data ($OHD$) \cite{Yu/2018,Moresco/2015} and Type Ia supernovae ($SNe\, Ia$) \cite{Scolnic/2018}. Here, we use the Pantheon sample for $SNe\,Ia$ data, which includes of 1048 points from the Low-z, SDSS, Pan-STARSS1 (PS1) Medium Deep Survey, SNLS, and HST surveys \cite{Chang/2019}.

\subsection{Hubble data}
We first employ a standard compilation of 31 measurements of Hubble data obtained through differential age method (DA). This method can be used to estimate the expansion rate of the universe at redshift $z$. Here, $H(z)$ can be calculated as $H(z)=-\frac{dz/dt}{(1+z)}$. To perform this analysis, we minimize 
\begin{equation}
    \chi^{2}_{H}= \sum_{i=1}^{31} \frac{\left[H(z_{i}, \mathcal{P})-H_{obs}(z_{i})\right]^{2} }{\sigma(z_{i})^{2}}.
\end{equation}
Here, $H(z_{i}, \mathcal{P})$ is the theoretical value for a given model at redshifts $z_{i}$, and $\mathcal{P}$ is the parameter space. $H_{obs}(z_{i})$ and $\sigma(z_{i})^2$ represents the observational value and error, respectively. 

\begin{widetext}

\begin{figure}[H]
\centering
\includegraphics[scale=0.5]{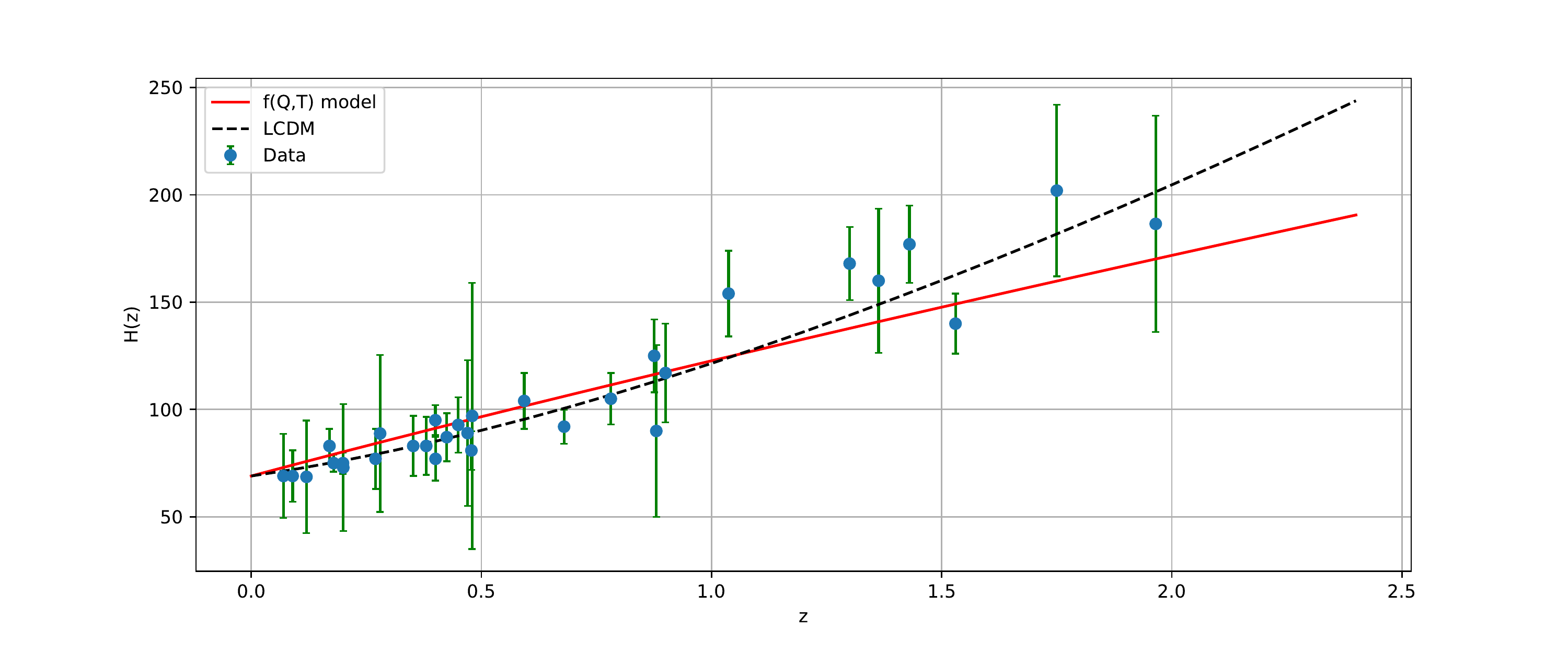}
\caption{Error bar plot of $H$ versus $z$ for $f(Q,T)$ model (red curve) and the $\Lambda$CDM model (black dotted curve). The blue dots depict the 31 points of the Hubble data}
\label{figure 1}
\end{figure}

\begin{figure}[H]
\centering
\includegraphics[scale=0.7]{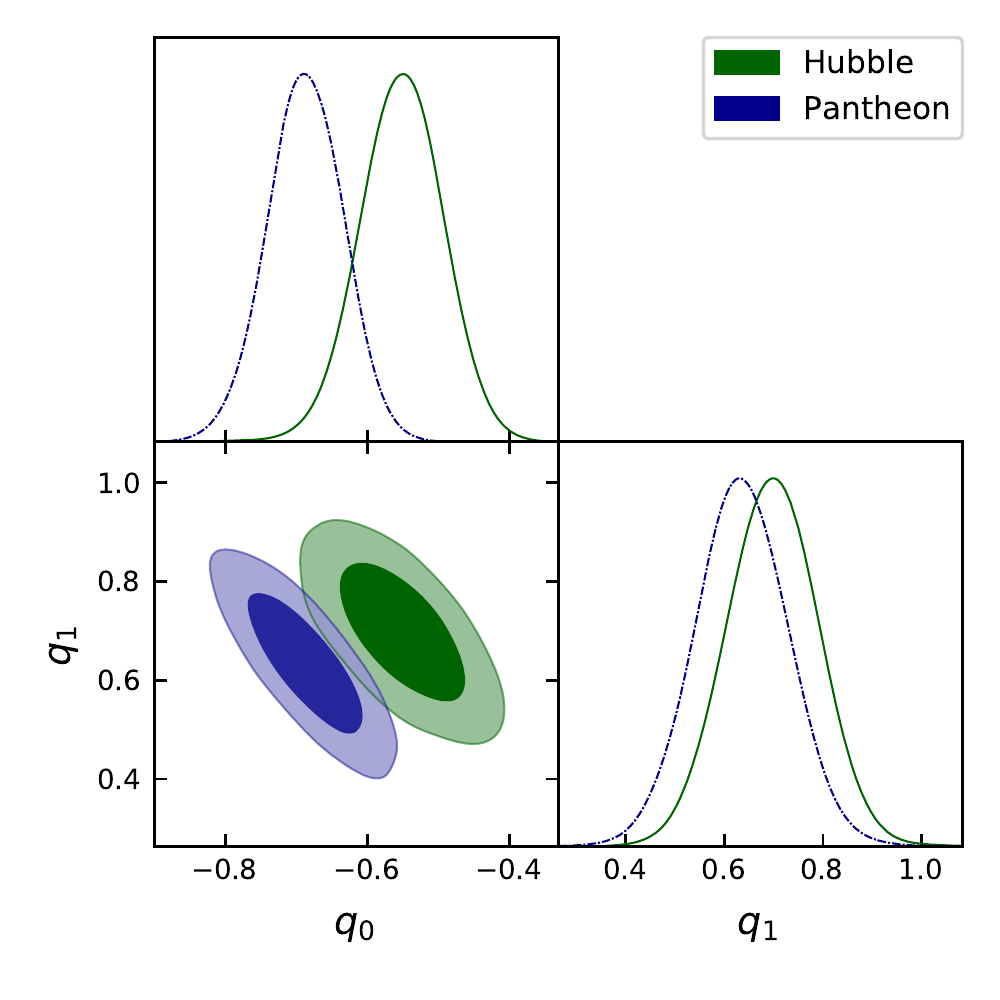}
\caption{The $1-\sigma$ and $2-\sigma$ confidence regions for the parameters corresponding to the Hubble and Pantheon data.}
\label{figure 3}
\end{figure}

\end{widetext}

\subsection{Pantheon data}
The Pantheon sample is a compilation of different supernova surveys that identified $SNe\,Ia$ at both high and low redshift. As a result, the whole sample spans the redshift range $0.01<z<2.26$. Provided the Tripp estimator \cite{Tripp/1998} with the light curve fitter, the normalized observational $SNe\,Ia$ distance modulus is given by $\mu= m_{B}-M_{B}+\alpha x_{1} -\beta c+\Delta_{M}+\Delta_{B}$, where $m_{B}$, $M_{B}$, and $c$ are the observed peak magnitude (at B-band maximum), absolute magnitude, and $SNe\,Ia$ color, respectively. Also, $\alpha$ and $\beta$ represent the relation between luminosity stretch and luminosity color, respectively. Further, the distance corrections on host galaxy mass and  simulation based expected biases are given by $\Delta_{M}$  and $\Delta_{B}$. \\
Based on accurate $SNe\,Ia$ simulations, Scolnic et al. \cite{Scolnic/2018,Kessler/2017} used the BEAMS with Bias Corrections (BBC) method to account for errors due to intrinsic scatter and selection effects. Hence, the distance modulus reduces to $\mu= m_{B}-M_{B}$. The expressions used for our analysis are as follows:
\begin{eqnarray}
\mu^{th} &=& 5log_{10}\left(\frac{d_{L}(z)}{1 Mpc}\right)+25,\\
d_{L}(z)&=& c(1+z) \int_{0}^{z} \frac{dz'}{H(z',\mathcal{P})},\\
\chi^{2}_{SN}&=& min \sum_{i,j=1}^{1048} \Delta \mu_{i} (C^{-1}_{SN})_{ij} \Delta \mu_{j}.
\end{eqnarray}

\begin{widetext}

\begin{figure}[H]
\centering
\includegraphics[scale=0.5]{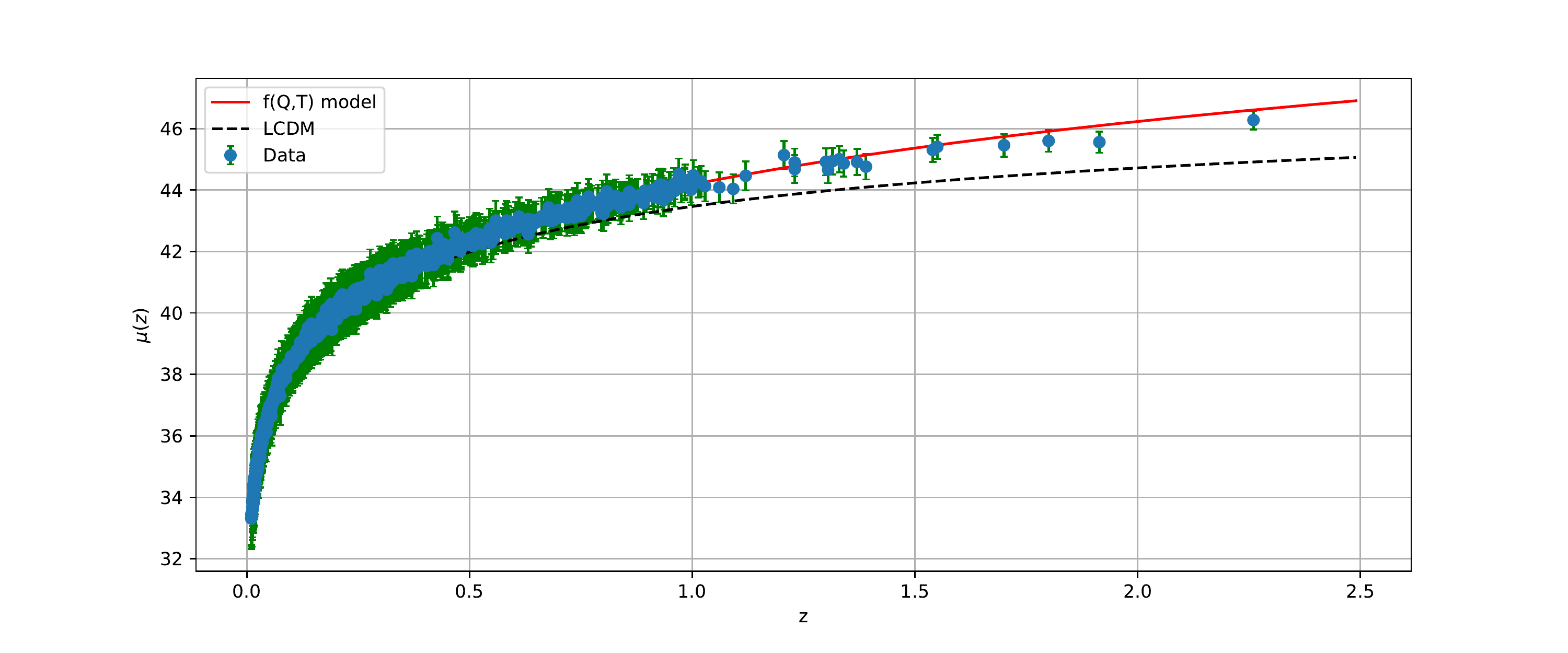}
\caption{Error bar plot of $\mu$ versus $z$ for $f(Q,T)$ model (red curve) and the $\Lambda$CDM model (black dotted curve). The blue dots depict the 1048 points of the Pantheon data}
\label{figure 2}
\end{figure}

\begin{figure}[H]
\centering
\includegraphics[scale=0.75]{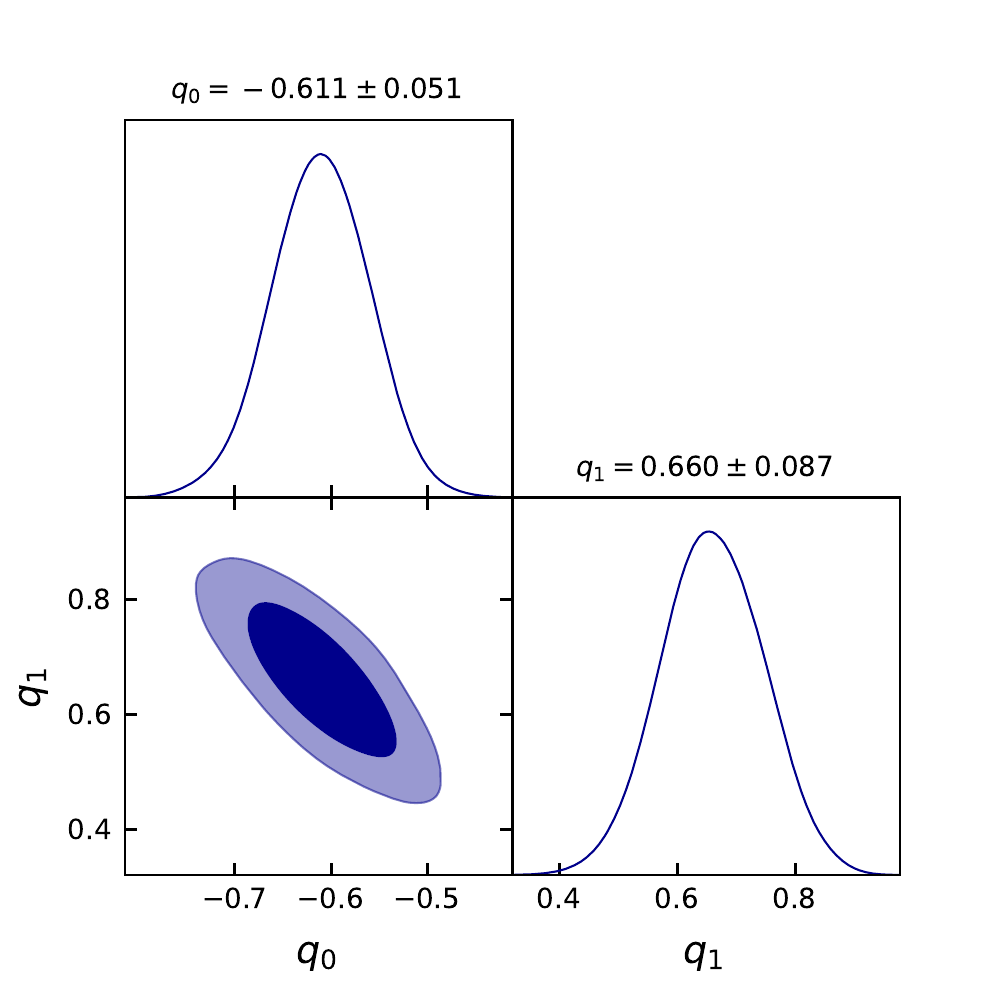}
\caption{The $1-\sigma$ and $2-\sigma$ confidence regions for the parameters corresponding to the joint data.}
\label{figure 4}
\end{figure}

\end{widetext}

\subsection{Hz+Pantheon}
Furthermore, we use the total likelihood function to get joint constraints for the parameters $q_{0}$ and $q_{1}$ from the Hubble and Pantheon samples. Henceforth, the relevant likelihood and Chi-square functions are given by
\begin{eqnarray}
\mathcal{L}_{joint} &=& \mathcal{L}_{OHD} \times \mathcal{L}_{SNeIa},\\
\chi^{2}_{joint} &=& \chi^{2}_{OHD} + \chi^{2}_{SNeIa}.
\end{eqnarray}
The constraints on model parameters are obtained by minimizing the respective $\chi^{2}$ using Markov Chain Monte Carlo (MCMC) and $emcee$ library. The results are obtained in Table I.
Further, figs. \ref{figure 1} and \ref{figure 2} shows the error bar fit for the considered model and the $\Lambda$CDM with $\Omega_{\Lambda_{0}}=0.7$, $\Omega_{m_{0}}=0.3$, and $H_{0}=69$ km/s/Mpc. The $1-\sigma$ and $2-\sigma$ contour plots are shown for Hubble, Pantheon and the joint observational data in figs. \ref{figure 3} and \ref{figure 4}.

\begin{widetext}

\begin{table}[h!]
\begin{center}
  \caption{Best-fit values of model parameters obtained from observational datasets}
    \label{table1}

\begin{tabular}{|l|c|c|c|c|}
\hline 
Datasets              & $q_0$ & $q_1$ & $z_{t}$ & $\omega_{0}$\\
\hline
$OHD$             & $-0.553^{+0.059}_{-0.059}$ \cite{Gadbail/2021a,Mamon/2017} & $0.698^{+0.092}_{-0.092}$ & $0.69^{+0.31}_{-0.17}$ \cite{Jesus/2020} & $-0.81^{+0.04}_{-0.04}$ \cite{Almada/2019,Mandal/2021}\\ 
\hline
$SNeIa$           & $-0.688^{+0.054}_{-0.054}$ \cite{Basilakos/2012,Almada/2019} & $0.635^{+0.094}_{-0.094}$ & $1.2^{+0.1}_{-0.22}$ \cite{Mamon/2017} & $-0.91^{+0.04}_{-0.04}$ \cite{Zhang/2010} \\
\hline
$OHD+SNeIa$     & $-0.611^{+0.051}_{-0.051}$ \cite{Gadbail/2021a} & $0.66^{+0.087}_{-0.087}$  & $0.87^{+0.15}_{-0.23}$ \cite{Mamon/2018,Farooq/2017} &  $-0.85^{+0.03}_{-0.04}$ \cite{Almada/2019,Mandal/2021}\\
\hline
\end{tabular}
\end{center}
\end{table}

\end{widetext}

\section{cosmological parameters}
\label{section 5}
\begin{figure}[H]
\includegraphics[scale=0.7]{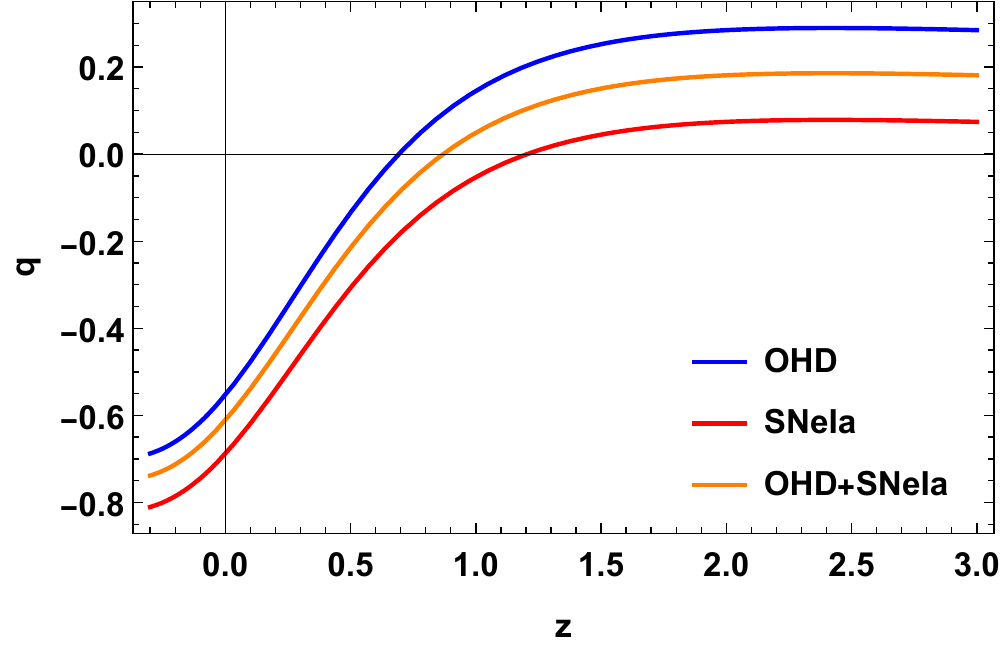}
\caption{The redshift evolution of the deceleration parameter.}
\label{figure 5}
\end{figure}

According to cosmological observational data, cosmic acceleration is a recent phenomenon. To capture the entire evolutionary history of the universe, a cosmological model must include both the decelerated and accelerated phases of the expansion. Hence, it is essential to investigate the behavior of the deceleration parameter $q$.\\
The behavior of $q$ for the associated values of model parameters constrained by the $OHD$, $SNe\,Ia$, and $OHD+SNe\,Ia$ is shown in fig. \ref{figure 5}. It is clear that the parameter $q$ shows a transition from a decelerated to an accelerated phase at redshift $z_{t}$. Further, the value of $z_{t}$ fluctuates in the range $0.5 - 1.3$ shown by recent observations \cite{Mehrabi/2021}.\\
Further, from figs. \ref{figure 6} and \ref{figure 7}, it is clear that the density parameter $\rho$ shows the positive behavior, whereas the pressure $p$ evolves negatively, respectively.

The equation of state parameter can indeed helpful in categorizing the various epochs of accelerated and decelerated expansion of the universe. The different epochs as follows: a) $\omega= 1$ represents a stiff fluid; b) $\omega= 1/3$ depicts the radiation-dominated phase; and c) $\omega= 0$ indicates the matter-dominated phase. The effective EoS parameter is defined as $\omega=\frac{p}{\rho}$.
Using Eqs. \eqref{29} and \eqref{30} in the above relation, we obtain $\omega$ in terms of $z$ as
 \begin{equation}
\label{32}
\omega=\frac{k_2 \left(q_0 \left(z^2+1\right)+q_1 (z+1) z+z^2+1\right)-k_1 \left(z^2+1\right)}{k_3 \left(z^2+1\right)+k_4 \left(q_0 \left(z^2+1\right)+q_1 (z+1) z+z^2+1\right)}
\end{equation}

\begin{figure}[H]
\includegraphics[scale=0.7]{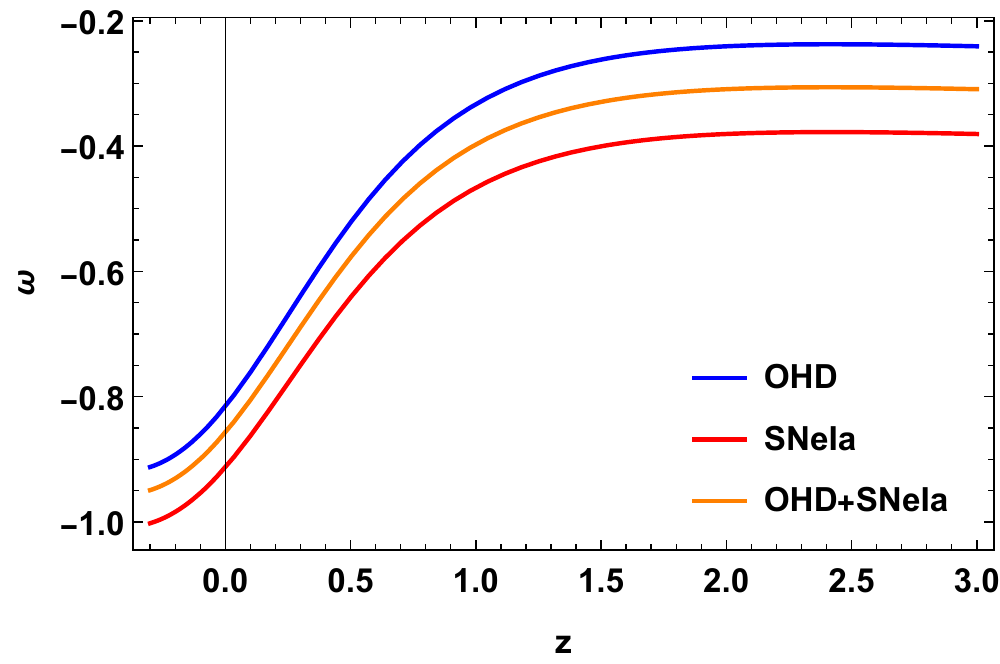}
\caption{The behavior of the EoS parameter $\omega$ versus redshift $z$.}
\label{figure 8}
\end{figure}

The evolution trajectory of the parameter $\omega$ is shown in fig. \eqref{figure 8}. One of the three possible states for the expanding universe is the cosmological constant $(\omega = -1)$, quintessence $(-1< \omega < -1/3),$ or phantom era $(\omega < -1)$. It is clear from fig. \ref{figure 8} that $\omega<0$ and shows a quintessence dark energy, which indicates an accelerating phase. In our model, it may be noted that the EoS parameter do not cross the phantom divide $\omega = -1$.
 
\begin{figure}[H]
\includegraphics[scale=0.7]{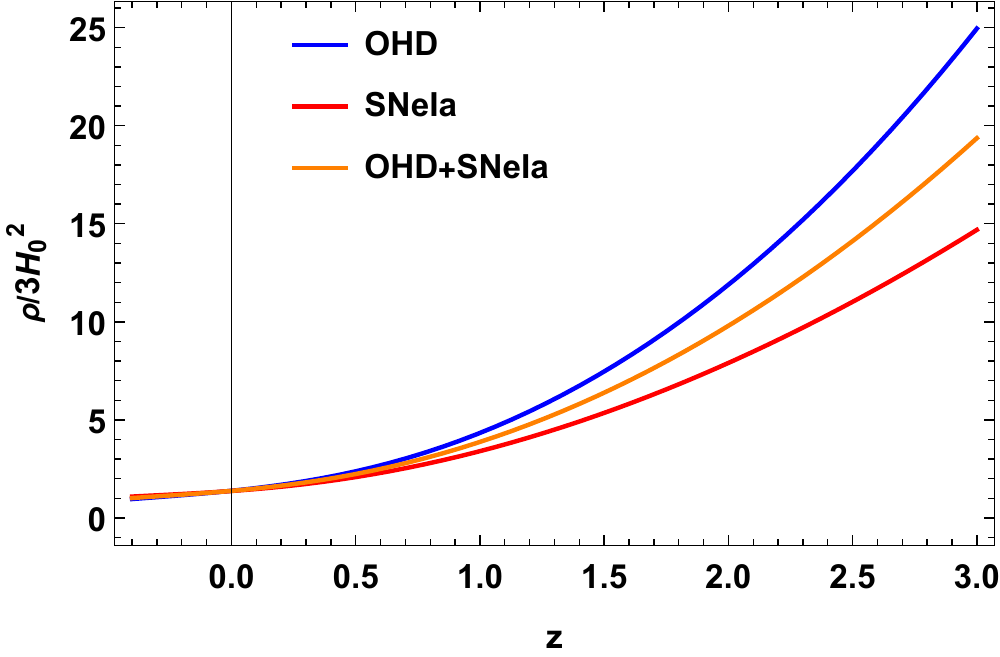}
\caption{The trajectory of the density parameter $\rho$ versus redshift $z$.}
\label{figure 6}
\end{figure}

\begin{figure}[H]
\includegraphics[scale=0.8]{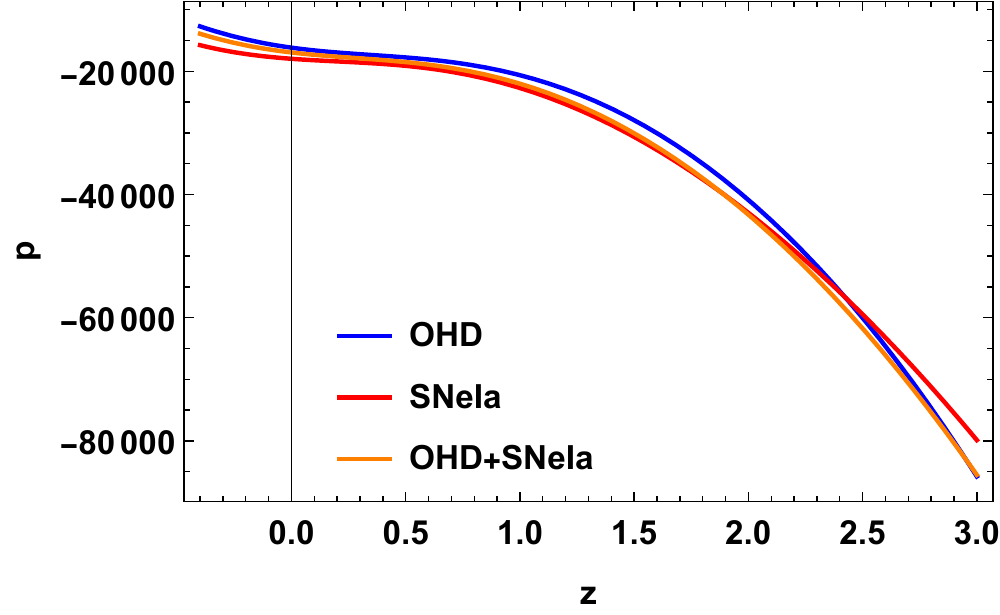}
\caption{The trajectory of the pressure $p$ versus redshift $z$.}
\label{figure 7}
\end{figure}

\section{Conclusion}
\label{section 6}
With time, the present scenario of the accelerating expansion of the universe has become even more fascinating. Numerous dynamical DE models and modified theories of gravity have been used to address this issue in various ways. Still, there is a search of suitable description of the accelerating universe. In the present work, we attempt to describe dark energy by assuming the parametric form of the deceleration parameter in the recently proposed Weyl-type $f(Q,T)$ gravity, where the non-metricity $Q$ is coupled to the trace $T$ of the energy-momentum tensor. From a geometric perspective, the Weyl type $f(Q,T)$ gravity may illustrate an alternative to dark energy and even dark matter.\\
Hence, we started with the simplest functional form $f(Q,T)=\alpha Q+\frac{\beta}{6\kappa^2}T$, where $\alpha$ and $\beta$ are model parameters, and the parametric form of deceleration parameter $q(z) = q_{0}+q_{1}\frac{z(1+z)}{1+z^2}$ as a function of $z$, where $q_0$ and $q_1$ are constants. With the help of Eq. \eqref{26}, we find the Hubble parameter in terms of redshift $z$. Furthermore, we constrained the parameters $q_0$ and $q_1$ by the MCMC technique using the $OHD$, $SNe\,Ia$, and $OHD+SNe\,Ia$ datasets (mentioned in table I). We have obtained the best-fit values for the $q_0$ and $q_1$ parameters in figs.  \ref{figure 3} and \ref{figure 4}. Since, the model parameters $\alpha$ and $\beta$ are not explicitly present in the expression of the Hubble parameter, we attempt to fix them to study the evolution of  density, pressure, and effective EoS parameters. We considered the values as $\alpha = -1.07$ and $\beta= 0.88$ from the reference \cite{Gadbail/2021}, that are constrained using the observational datasets.

Finally, it is seen that the deceleration parameter $q(z)$ smoothly switches from a decelerated to an accelerated phase of expansion. Also, in figs. \ref{figure 6} and \ref{figure 7}, the energy density parameter decreases as the universe continues to expand in the far future, and pressure increases in the negative behavior. The effective EoS parameter exhibits the negative behavior indicating the accelerating universe and the quintessence dark energy.
However, the present values of the deceleration and effective EoS parameters are more compatible with the most recent values of cosmological parameters observed for $OHD$ data. It is work mentioning that the solution of the Hubble parameter obtained from a divergence-free parametric form of deceleration parameter is in good agreement with Weyl-type $f(Q,T)$ gravity. 

\section*{Data Availability Statement}
There are no new data associated with this article.

\section*{Acknowledgments}
GNG acknowledges University Grants Commission (UGC), New Delhi, India for awarding Junior Research Fellowship (UGC-Ref. No.: 201610122060). SA acknowledges CSIR, Govt. of India, New Delhi, for awarding Senior Research Fellowship. PKS acknowledges CSIR, New Delhi, India for financial support to carry out the Research project [No.03(1454)/19/EMR-II Dt.02/08/2019].

\end{document}